\documentclass[twocolumn,showpacs,preprintnumbers,floatfix,apl]{revtex4}

\usepackage{graphicx}
\usepackage{subfigure}
\usepackage[]{amssymb}
\usepackage[]{amsmath}

\begin{document}

\title{Controlling the partial coalescence of a droplet on a vertically vibrated bath}

\author{T. Gilet, N. Vandewalle and S. Dorbolo}

\affiliation{GRASP, Physics Department, University of Li\`{e}ge, B-4000 Li\`{e}ge, Belgium.}

\begin{abstract}
A new method is proposed to stop the cascade of partial coalescences of a droplet laid on a liquid bath. The
strategy consists in vibrating the bath in the vertical direction in order to keep small droplets bouncing.
Since large droplets are not able to bounce, they partially coalesce until they reach a critical size. The
system behaves as a low pass filter : droplets smaller than the critical size are selected. This size has been
investigated as a function of the acceleration and the frequency of the bath vibration. Results suggest that the
limit size for bouncing is related to the first mode of the droplet deformation.
\end{abstract}

\maketitle

\section{Introduction}

Liquid droplets are studied more and more intensively in the framework of microfluidic applications
\cite{Stone:2004}. Among possible applications, droplets can be used in chemical engineering in order to mix
tiny amounts of reactive substances. To achieve such a goal, one has to invent some processes to manipulate a
droplet : motion, binary collision, fragmentation... To prevent droplets from contamination, it is also
important to avoid any contact with a solid element. This paper introduces a method that could eventually
realize all those operations. The basic idea is to combine two physical phenomena : the delayed coalescence on a
vibrated interface, and the partial coalescence.

First, we describe both phenomena separately and identify the physical conditions required to achieve these
effects.

\subsection{Partial coalescence}

\begin{figure}[htbp]
\includegraphics[width=\columnwidth] {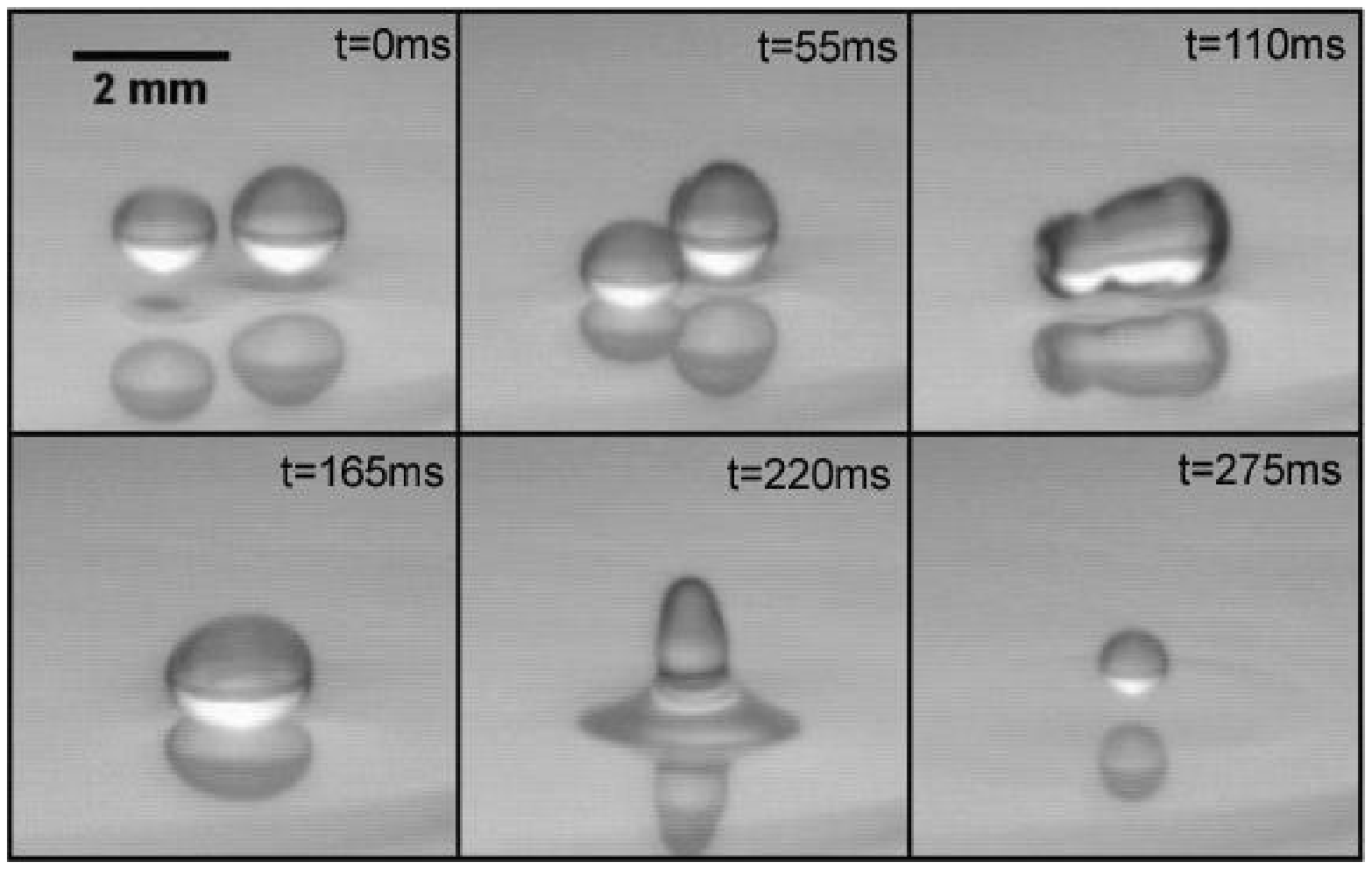}
\caption{\label{fig:Montage} Two droplets are bouncing on a vibrated liquid/air interface. They meet and
coalesce. The new droplet is too large to take off from the interface and to bounce, so it eventually coalesces
partially after 110 ms. The daughter droplet is small enough to bounce again for several tens of seconds. The
six pictures in this sequence are taken every 55 ms.}
\end{figure}

When a droplet is gently laid on a liquid bath at rest, it very quickly coalesces with the bath. Because of the
low air viscosity, the drainage of the air film between the droplet and the bath lasts less than a second (a
similar drainage process can be found in antibubbles \cite{Dorbolo:2005}). Suddenly, the film becomes thin
enough to break and there is a contact between the droplet and the bath.  A droplet made of a low-viscosity
liquid can experience a partial coalescence \cite{Vandewalle:2006}: it does not fully empty, and a new smaller
droplet is formed above the bath interface (as illustrated in the second image line of Fig.\ref{fig:Montage}).
This daughter droplet can also coalesce partially. The process is repeated until the droplet becomes small
enough to coalesce totally. When the droplet is made of a non-viscous liquid surrounded by air, the daughter
droplet has a radius $R_{i+1}$ that is about half \cite{Gilet:2007} of the mother one ($R_i$) :
\begin{equation} \label{eq:PartCoal}
\frac{R_{i+1}}{R_i} \simeq 0.5
\end{equation}
This phenomenon was first reported in 1930 by Mahajan \cite{Mahajan:1930}, and investigated by Charles and Mason
in 1960 \cite{Charles:1960}. In their experiments, another immiscible liquid was surrounding the droplet
(instead of air). Recently, many studies have focused on understanding this phenomenon
\cite{Gilet:2007,Leblanc:1993,Thoroddsen:2000,Blanchette:2006,Chen2:2006,Aryafar:2006}. In general, gravity and
viscosity forces in both fluids tend to favor total coalescence \cite{Gilet:2007}. Partial coalescence is only
possible when surface tension is the dominant force. A measure of the relative influence of the droplet
viscosity compared to surface tension is given by the Ohnesorge number
\begin{equation} \label{eq:Ohnesorge}
Oh = \nu \sqrt{\frac{\rho}{\sigma R_i}}
\end{equation}
where $\nu$ is the viscosity of the liquid, $\rho$ its density, and $\sigma$ the surface tension between the
droplet and the air. As shown in \cite{Leblanc:1993,Blanchette:2006,Gilet:2007}, the critical Ohnesorge
corresponding to the transition from partial to total coalescence is approximately $Oh_c \simeq 0.026$. Small
droplets ($Oh>Oh_c$) undergo a total coalescence while larger ones ($Oh < Oh_c$) coalesce only partially.

\subsection{Delayed coalescence}

Typically, droplets cannot stay more than a second on the liquid/air interface, due to gravity. In 2005, Couder
\textit{et al.} \cite{Couder:2005,CouderNat:2005} have found an original way to avoid coalescence. In their
experiment, the liquid bath is vertically vibrated by an eletromagnetic shaker. The sinusoidal vibration has a
pulsation $\omega$ and an amplitude $A$. When the reduced acceleration $\Gamma = A \omega^2 / g$ is sufficiently
large, the droplet bounces periodically on the vibrated interface. The minimum acceleration $\Gamma_m$ for
bouncing to occur has been estimated by Couder \textit{et al.} \cite{Couder:2005} in the case of viscous
droplets \footnote{To obtain Eq.(\ref{eq:Couder}), Couder \textit{et al.} assume that the droplet is not able to
store interfacial energy during the bouncing. The excess of kinetic energy at impact is then dissipated by
viscous forces.} as:
\begin{equation} \label{eq:Couder}
\Gamma_m = 1 + \frac{1}{\mathrm{Re_c}} \frac{\rho_a}{\rho_l} \frac{\omega^2}{g} \frac{r_R^4}{R^3}
\end{equation}
where $\mathrm{Re_c}$ is a critical Reynolds number (constant), $\rho_a$ the air density, $\rho_l$ the liquid
density, $g$ the gravity acceleration, $R$ the droplet radius (when spherical) and $r_R$ the horizontal
extension of the film between the droplet and the bath. Although $r_R$ has no simple analytical form, it has
been shown \cite{Couder:2005} that $\Gamma_m$ is monotonically increasing with $R$. This means that for a given
reduced acceleration $\Gamma$, there is a critical size $R_M(\Gamma)$ such as
\begin{equation} \label{eq:BouncingCondition}
\begin{array}{lll}
R < R_M & \rightarrow \mbox{ Bouncing } & \rightarrow \mbox{ Stabilization } \\
R > R_M & \rightarrow \mbox{ No bouncing }& \rightarrow \mbox{ Coalescence }
\end{array}
\end{equation}
One can reasonably think that this conclusion is still valid when droplets are not viscous. This point will be
shown below. In reality, droplets experience a finite lifetime before coalescence, even when they bounce: the
film between the droplet and the bath breaks and the droplet coalesces \cite{Terwagne:2007}, sometimes after a
thousand bounces. Moreover, when the vibration amplitude is too high, steady waves appear on the bath interface,
due to the Faraday instability \cite{Benjamin:1954}. Under these conditions, the trajectory of the droplet
generally becomes chaotic and the lifetime is considerably reduced. The more viscous the bath is, the higher is
the Faraday acceleration threshold $\Gamma_F$. Couder \textit{et al.} have noted that when the bath is vibrated
just below $\Gamma_F$, the droplets are sometimes able to move horizontally on the interface
\cite{CouderNat:2005,CouderJPCM:2005}. In this regime, droplets can strongly interact among themselves or with
obstacles \cite{CouderPRL:2006}.

In the present study, the key idea is to combine both effects: (i) the partial coalescence, that allows to
change the radius of the droplet according to Eq.(\ref{eq:PartCoal}) and (ii) the delayed coalescence that
becomes efficient when $R_{i+1} < R_M$. Conditions on surface tension and viscosity have to be satisfied in
order to observe both phenomena. According to the Ohnesorge criterion \footnote{The Ohnesorge criterion is
designed for interfaces at rest. However, it seems to prevail for interfaces in motion : only the critical
Ohnesorge $Oh_c$ should be slightly different.}, the kinematic viscosity of the droplet liquid has to be lower
than 3.7cSt in order to ensure the partial coalescence of a millimetric droplet (with $\rho_l \simeq
1000$kg/m$^3$ and $\sigma \simeq 20 \times 10^{-3}$N/m). For such a low viscosity, $\Gamma_F \ll 1$ while
$\Gamma_m > 1$ according to Eq.(\ref{eq:Couder}). Therefore, it does not seem possible to stop a cascade of
partial coalescences when using the same viscosity for the droplet and for the bath. However, one can satisfy
Eq.(\ref{eq:Ohnesorge}) and Eq.(\ref{eq:BouncingCondition}) by considering a droplet made of silicon oil with a
low viscosity (0.65cSt) bouncing on a bath with a large viscosity (1000cSt).

\section{Experimental set-up}

We apply vertical vibrations to a 7 mm-thick bath of 1000cSt silicon oil in order to recover the daughter
droplets of the partial coalescence. The frequency of the bath vibration is varied from 40 to 100Hz. The
acceleration is tuned from 0 to 2.5g and measured with an accelerometer; the precision is estimated below
0.4m/s$^2$, $g$ being the gravity.

During a partial coalescence, the coalescing part of the droplet forms a pool at the bath surface. The liquid of
this pool cannot immediately mix with the surrounding highly viscous oil. Consequently, the daughter droplet of
the partial coalescence has unavoidably to make its firsts bounces on the pool formed by its own mother droplet.
This does not perturbe the bouncing too much. Sometimes, due to inertial effects, the daughter droplet does not
succeed in quickly stabilizing its bouncing on the vibrating interface (it was observed even when $R<R_M$). In
this case, another partial coalescence is necessary.

A fast video recorder (Redlake Motion Pro) is placed near the surface. Movies of coalescences have been recorded
at a rate of up to 1000 frames per second. The droplet diameter is measured from the images with an absolute
error due to the finite pixel size (about 30$\mu$m). Since the droplet is constantly deforming, this measurement
is made on the few images where the droplet appears the most spherical.

\section{Results}

Figure \ref{fig:Montage} illustrates a typical scenario observed with droplets on a vibrated interface. Two
small droplets are laid next to each other on the bath. They are small enough to take off and bounce. They
approach each other and coalesce into a large droplet. This resulting droplet is now too large to take off, as
the thin air film that separates it from the bath cannot be regenerated during the vibration. The droplet
experiences a few oscillations before coalescing: it is unstable. The coalescence is partial and the daughter
droplet is small enough to bounce again. The system behaves as a low pass filter: partial coalescences occur,
decreasing the radius of the droplet, until the radius of the droplet respects the condition expressed in
Eq.(\ref{eq:BouncingCondition}). This experiment is repeated several times, with various droplet sizes.

\begin{figure}[htbp]
\includegraphics[width=\columnwidth] {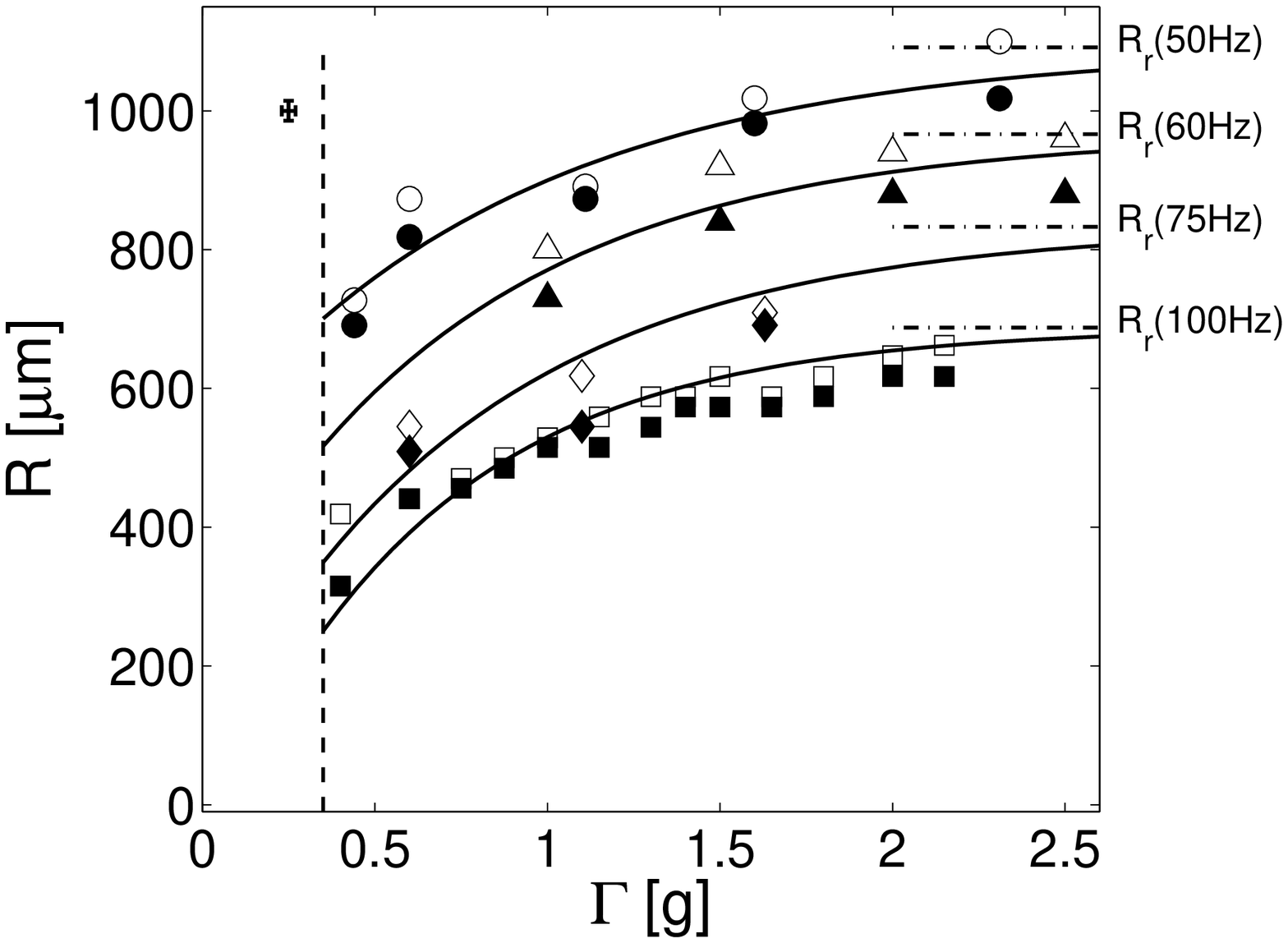}
\caption{\label{fig:PhaseDiag} Phase diagram ($\Gamma$-$R$). Critical radius $R_M$ as a function of the bath
reduced acceleration $\Gamma = A \omega^2/g$ for various frequencies : 50Hz($\bullet$) - 60Hz($\blacktriangle$)
- 75Hz($\blacklozenge$) - 100Hz($\blacksquare$). Black symbols correspond to the largest observed stable
droplets while open symbols correspond to the smallest observed unstable droplets. The critical radius
$R_M(\Gamma,f)$ is located between these two bounds. The dashed vertical line is the minimum acceleration for
bouncing. The dashdot horizontal line correspond to the asymptotic size $R_a(f)$, the maximum size of stable
droplets, whatever the acceleration. This size approximately correspond to $R_r$, given by Eq.(\ref{eq:Courty}).
The typical error bar is illustrated in the upper left corner of the figure.}
\end{figure}

The radii $R$ of the smallest unstable (open symbols) and the largest stable (black symbols) observed droplets
are plotted as a function of the bath acceleration $\Gamma$ in Fig.\ref{fig:PhaseDiag}, for various frequencies
50, 60, 75 and 100Hz (see legend).  Between both limit values of $R$, a mean critical value $R_M(\Gamma,f)$ may
be defined as the maximum value of the radius for a stable bouncing droplet.

The minimum acceleration $\Gamma_{min}=0.35$ needed to stabilize droplets is lower than 1. It seems impossible
to find a bouncing droplet on the left of the vertical dashed line. The on-off behavior of the stabilization
process around $\Gamma_{min}$ has to be nuanced. Indeed, the lifetime of a bouncing droplet should be considered
close to the boundaries. Some combinations of the parameters probably stabilize droplets for a longer time than
others, or allow to stop the cascade of partial coalescences at an intermediate value between $R_M = 0$ and $R_M
= 0.4$mm.

\section{Discussion}

In our experiments, low viscous droplets are bouncing on a high viscous bath. It means that the deformation of
the droplet is larger than the deformation of the bath : the droplet is more able to store and restore
interfacial energy in order to ensure the bouncing. Conditions are comparable to an elastic ball bouncing on a
rigid plate.

First, according to Eq.(\ref{eq:Couder}), $\Gamma_{min}=1$ for viscous droplets. However, since non-viscous
droplets can restore surface energy (the dissipation being limited), the bouncing is possible for reduced
accelerations lower than 1: A restitution coefficient has to be taken into account. The analogy between the
non-viscous droplet and the elastic ball (defined and analyzed in \cite{Luck:1993}) allows us to define this
restitution coefficient $\alpha$ of the droplet, such as:
\begin{equation}
\Gamma_{min} = \frac{1-\alpha}{1+\alpha} \simeq 0.35 \rightarrow \alpha \simeq 0.48
\end{equation}
This value is higher than the restitution coefficient of 0.22 found in previous works
\cite{Jayaratne:1964,Honey:2006}. However, some aspects have to be taken into account : (1) In those
experiments, small (less deformable) droplets bounce on a low-viscous (deformable) bath. The restitution
mechanism is consequently not the same as in our case. (2) Moreover, for elastic beads, it is known that the
restitution coefficient tends to 1 when the impact velocity tends to 0 \cite{Gerl:1999}. In our case, impact
velocities near the critical acceleration are about $v = \Gamma g / \omega \leq 2$cm/s while in
\cite{Honey:2006}, impact velocities are at least ten times higher. This could also explain the difference
between restitution coefficients in both configurations.

Second, experimental points of Fig.\ref{fig:PhaseDiag} define two regions in the $\Gamma-R$ diagram.  The
boundaries that depend on the frequency are shown as continuous curves based on a qualitative fit. For a given
frequency, droplets located above the curve are unstable while droplets located below are stable. More
precisely, when a droplet is created with a radius larger than the critical droplet radius $R_M(\Gamma, f)$, the
droplet partially coalesces until it reaches the stable region (below the curves). The stability threshold
$R_M(\Gamma,f)$ slightly increases with the forcing acceleration, as in Eq.(\ref{eq:Couder}). The curves seem to
saturate at high accelerations towards an asymptotical value $R_a(f)$ that decreases with an increasing
frequency.

Since the bath is much more viscous than the droplet, the droplet deformation is the main elastic mechanism able
to ensure the bouncing. A characteristic size related to the deformation is given by the wavelength of the first
normal mode. According to \cite{Courty:2006}, for a sitting droplet, this wavelength $\lambda$ satisfies to:
\begin{equation} \label{eq:Courty}
\lambda^3 = (\pi R_r)^3 = \frac{2 \pi \sigma}{\rho f^2 \biggl( 1 + \sqrt{\frac{5}{4\pi}}\biggr)}
\end{equation}
The radius $R_r$ corresponding to $\lambda$ is represented in Fig.\ref{fig:PhaseDiag}. It fits well to the
asymptotical value $R_a$ for high accelerations. This fact suggests that the first mode of droplets deformation
is related to the maximum bouncing droplet size for a given frequency.

\begin{figure}[htbp]
\includegraphics[width=\columnwidth] {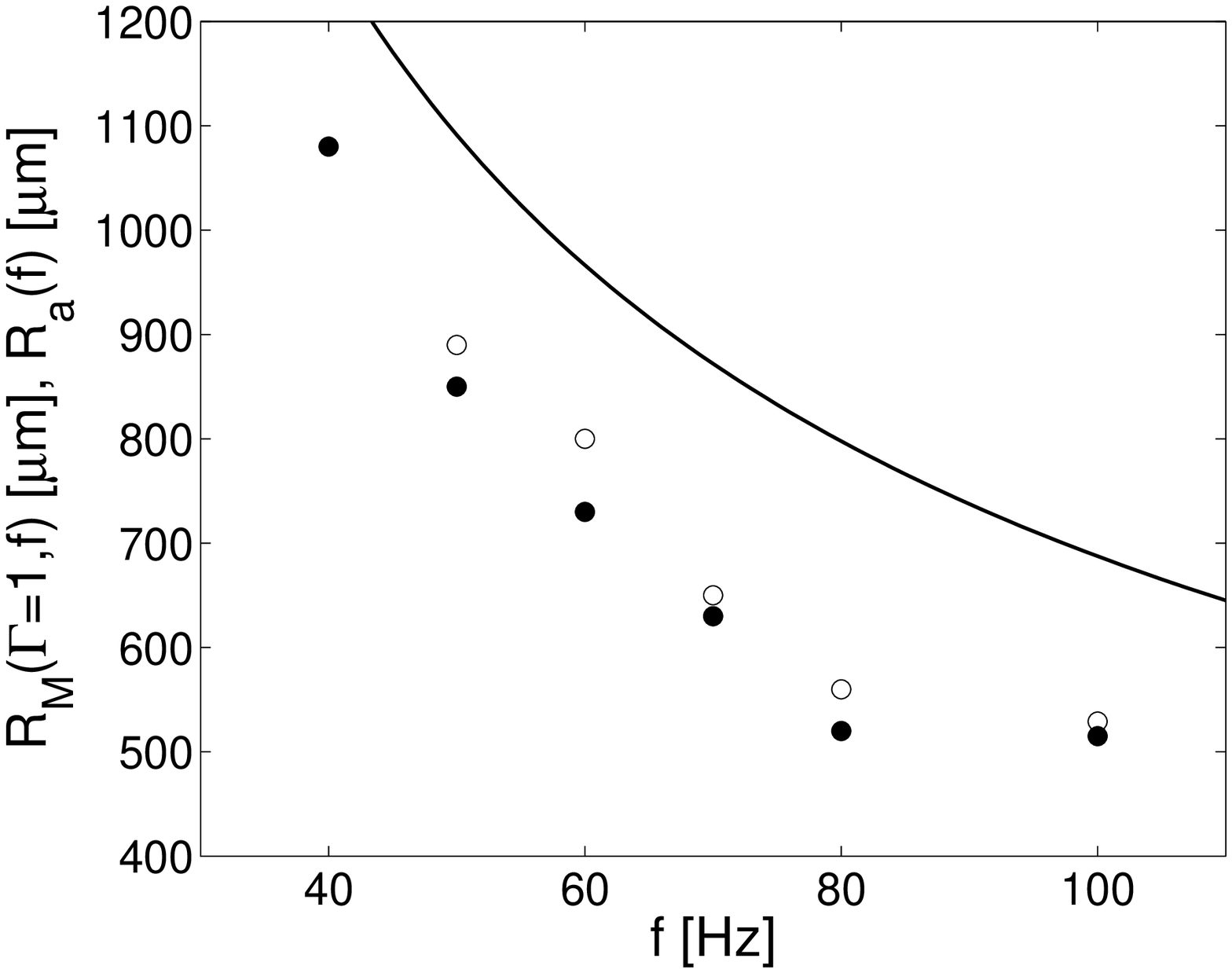}
\caption{\label{fig:PhaseDiag2} Critical radius $R_M$ as a function of the frequency when $\Gamma = 1$. Black
circles correspond to the largest observed stable droplets while open circles correspond to the smallest
observed unstable droplets. The critical radius $R_M$ is located between these two bounds. The continuous line
corresponds to Eq.(\ref{eq:Courty}).}
\end{figure}

Finally, the critical radius $R_M$ has been measured as a function of the frequency, the acceleration being
fixed $\Gamma=1$ (Fig.\ref{fig:PhaseDiag2}). Open circles correspond to the smallest unstable droplets, while
black circles are related to the largest stable ones. The continuous line represents Eq.(\ref{eq:Courty}). The
curve does not fit the experimental results because the asymptotic regime is not reached when $\Gamma=1$.
Moreover, sitting and bouncing droplet geometries are not exactly equivalent. However, we found that the
decreasing of $R_M$ with increasing frequencies (at $\Gamma=1$) is coherent with the decreasing of $R_r$.

\section{Conclusions}

In summary, we have developed a method to stop a cascade of partial coalescences and to maintain the residual
droplet alive on the interface. Our approach is based on the experiment of Couder \cite{Couder:2005}, that
consists in the bouncing of a droplet on a bath subjected to a vertical vibration. We have shown that it is
essential to use a viscous liquid for the bath and a low-viscous one for the droplet.  The threshold radius
below which the droplet can bounce has been found to depend on the acceleration and on the frequency of the
bath. By tuning these forcing parameters, it is possible to choose the maximum allowed size for bouncing
droplets in a broad range. A concordance is observed between the radius of the largest stable droplet (for a
given frequency) and the radius related to the first normal mode of deformation.

TG and SD thank FRIA/FNRS for financial support. Part of this work has been supported by Colgate-Palmolive. H.
Caps, D. Terwagne and A. Bourlioux are acknowledged for fruitful discussions.

\end{document}